\documentclass{aastex6}
\usepackage{mkfig}

\begin{document}

\title{A {\em Ulysses} Detection of Secondary Helium Neutrals}

\author{Brian E. Wood}
\affil{Naval Research Laboratory, Space Science Division,
  Washington, DC 20375, USA; brian.wood@nrl.navy.mil}

\author{Hans-Reinhard M\"{u}ller}
\affil{Department of Physics and Astronomy, Dartmouth College,
  Hanover, NH 03755, USA}

\author{Manfred Witte}
\affil{Max Planck Institute for Solar System Research,
  G\"{o}ttingen D-37077, Germany}


\begin{abstract}

The {\em Interstellar Boundary
EXplorer} (IBEX) mission has recently studied the flow of interstellar
neutral He atoms through the solar system, and discovered the existence
of a secondary He flow likely originating in the outer heliosheath.
We find evidence for this secondary component in {\em Ulysses} data.
By coadding hundreds of {\em Ulysses} He beam
maps together to maximize signal-to-noise, we identify a weak
signal that is credibly associated with the secondary component.  Assuming
a laminar flow from infinity, we infer the following He flow
parameters:  $V=12.8\pm 1.9$ km~s$^{-1}$,
$\lambda=74.4\pm 1.8^{\circ}$, $\beta=-10.5\pm 4.1^{\circ}$, and
$T=3000\pm 1100$~K; where $\lambda$ and $\beta$ are the ecliptic
longitude and latitude direction in J2000 coordinates.  The secondary
component has a density that is $4.9\pm 0.9$\% that of the primary component.
These measurements are reasonably consistent with measurements from
IBEX, with the exception of temperature, where our temperature is much
lower than IBEX's $T=9500$~K.  Even the higher IBEX temperature is
suspiciously low compared to expectactions for the outer heliosheath
source region.  The implausibly low temperatures are
due to the incorrect assumption of a laminar flow instead of
a diverging one, given that the flow in the outer heliosheath source
region will be deflecting around the heliopause.  As for
why the IBEX and {\em Ulysses} $T$ values are different, difficulties
with background subtraction in the {\em Ulysses} data are a potential
source of concern, but the discrepancy may also be another effect of
the improper laminar flow assumption, which could affect the IBEX and
{\em Ulysses} analyses differently.

\end{abstract}


\section{Introduction}

     The heliopause is the boundary separating plasma flow
associated with the solar wind and the plasma flow
of the interstellar medium (ISM) past the Sun.  However, the
local ISM is not fully ionized.  Both H and He are not only partially
but probably mostly neutral \citep{vi03}.  Unlike the
ions, neutrals can penetrate the heliopause.  It is possible,
therefore, to observe neutrals in the inner solar system that
are largely unaffected by their passage through the heliosphere, other
than by solar gravity and photoionization.  However, charge exchange
processes throughout the heliosphere create other populations of
neutrals as well, with properties that are diagnostic of the plasma
properties in the regions in which the charge exchange
occurred.  The {\em Interstellar Boundary EXplorer} (IBEX)
mission, launched in 2008, is designed to study these neutral
particles, as well as the pristine ISM flow \citep{djm09}.

     The ISM neutrals streaming through the inner solar system are
at the lowest energies accessible to IBEX.
A major goal of IBEX is to measure the properties of the
undisturbed ISM surrounding the Sun using observations of these
neutrals \citep{mb12,em12,djm15,jms15,nas16}.
Of particular interest are the neutral He atoms,
because unlike H, He has low charge exchange cross
sections, and the vast majority of ISM He that approaches the
heliosphere can reach the inner solar system without undergoing
any charge exchange interaction.  Thus, He is better suited for
studying the undisturbed ISM than H, despite an abundance that is
an order of magnitude below that of H.

     Studies of the low energy He flow observed by IBEX discovered that
there are two components to the flow, the primary component representing
the ISM He particles, and a second component termed the
``Warm Breeze'' \citep{mak14}, a component also later detected for
oxygen \citep{jp16}.  Charge exchange cross sections
involving He are low, but they are not zero, and subsequent
analysis strongly suggests that this second component is created by
charge exchange in the outer heliosheath just beyond the heliopause
\citep{mak16,mb17}.  The dominant charge exchange reaction is
${\rm He^{0}} + {\rm He}^{+}\rightarrow {\rm He}^{+} + {\rm He^{0}}$,
which is important due to the significant abundance of both ${\rm He^{0}}$
and ${\rm He}^{+}$ in the ISM \citep{mb12,hrm13}.

     Measurements of the ``Warm Breeze'' flow parameters have
so far relied on the same codes used to analyze
the primary component, assuming a laminar flow from infinity.  This
assumption leads to an inferred flow speed of $V=11.3$ km~s$^{-1}$ towards
ecliptic coordinates ($\lambda$,$\beta$)=($71.6^{\circ}$,$-12.0^{\circ}$),
with a temperature of $T=9500$~K, and an abundance at $5.7\%$ of the
primary ISM component \citep{mak16}.  The temperature is almost
certainly too low to be representative of the true temperature in
the outer heliosheath source region, where temperatures of
$T>10,000$~K with significant gradients are expected
\citep[e.g.,][]{gpz13,vvi15,mb17}.
This underestimation of $T$ is an effect of the divergence of
the He flow in the outer heliosheath.  A divergent flow will
narrow the velocity distribution.  Modeling this flow assuming a laminar
flow will fail to take this narrowing into account and will naturally
lead to underestimates of temperature \citep{bew17}.

     Prior to IBEX, the GAS instrument on board the long-lived
{\em Ulysses} mission studied the neutral He flow intermittently
during its 1990--2007 lifetime \citep{mw93,mw96,mw04}.
Although {\em Ulysses} cannot
match the high signal-to-noise (S/N) of IBEX,
{\em Ulysses} possesses advantages that make it worthwhile to
still consider the observational constraints that it can offer.  The
primary advantage is that {\em Ulysses} made observations at different
distances from the Sun and at locations below, above, and within the
ecliptic plane \citep{kpw92}; whereas IBEX makes observations of the He
flow from the same location in Earth's orbit around the Sun at the same
time every year \citep{mb12,em12,djm15}.  At least in
the analysis of the primary He flow component, the variation in
observation locale for {\em Ulysses} breaks parameter degeneracies
that plague the analysis of IBEX data, leading to tighter error bars
on the flow parameters despite the lower S/N \citep{bew15a,bew15b}.
Although no evidence of a secondary He flow has
been reported in past analyses of the {\em Ulysses} data, we here take
a closer look at the data to see if a signature of the secondary
component can be found, the goal being to see whether {\em Ulysses}
measurements can confirm the IBEX detection and if so, to see whether
{\em Ulysses} observational constraints are consistent with the IBEX
measurements of the flow parameters.

\section{Searching for Secondary Helium Neutrals}

     After launch in 1990 October and a gravitational assist from
Jupiter in 1992 February, {\em Ulysses} achieved its final
intended orbit nearly perpendicular to the ecliptic plane, with
an aphelion near Jupiter's distance of 5~AU and a perihelion
near 1~AU.  The GAS instrument on board {\em Ulysses} provided the
first direct in~situ measurements of interstellar neutral He atoms
in the inner heliosphere \citep{mw92}.  Detection of the
He particles could only happen when their inflow velocity was high
enough to exceed the particle energy detection threshold of the GAS
instrument, and this only occurred when the {\em Ulysses} spacecraft
was moving quickly in the part of its orbit closest to the Sun.
Thus, the He observations are confined to the three fast latitude
scans in 1994--1996, 2000--2002, and
2006--2007.  The GAS instrument works essentially like a pinhole
camera, which would gradually map the He beam on the sky by scanning
over it in a manner defined by the rotation axis of {\em Ulysses}.
A map would typically be completed over the course of $2-3$ days,
with the final {\em Ulysses} database consisting of $\sim 400$ maps
through GAS's wide field of view (WFOV) channel, and $\sim 400$
through its narrow field of view (NFOV) channel.

     The first analyses of the He data were made while the
{\em Ulysses} mission was still in operation \citep{mw92,mw93,mw96,mw04}.
More recent analyses are by \citet{oak14}, \citet{mb14}, and
\citet{bew15b}.  We reanalyzed the full {\em Ulysses} data set
\citep[][hereafter WMW15]{bew15b}, motivated in part by
initial discrepancies that seemed to exist between the IBEX He
measurements and the {\em Ulysses} ones \citep{mb12,em12,pcf13},
discrepancies which have since mostly been
resolved \citep{djm15}.  In our reanalysis, we confined
our attention to 238 WFOV maps that fully cover the He beam, and
are not plagued by obvious artifacts or large background gradients.
Our search for He secondary neutrals can be considered a follow-up
analysis to WMW15, as we will be using the same 238 He beam maps
described there.

     We refer the reader to Figure~1 of WMW15 for an example
of what an individual {\em Ulysses} beam map looks like.  Searching
for hints of the secondary neutrals in individual
maps like that would be very difficult for a couple
reasons.  The first is that the individual maps are unlikely to
have sufficient S/N to clearly detect the weak secondary signal,
and the second is that the individual maps only cover the part of
the sky surrounding the primary He beam, and will generally not
completely extend over the secondary He beam as well, making it
difficult to discern from the background,
regardless of S/N.

\begin{figure}[t]
\plotfiddle{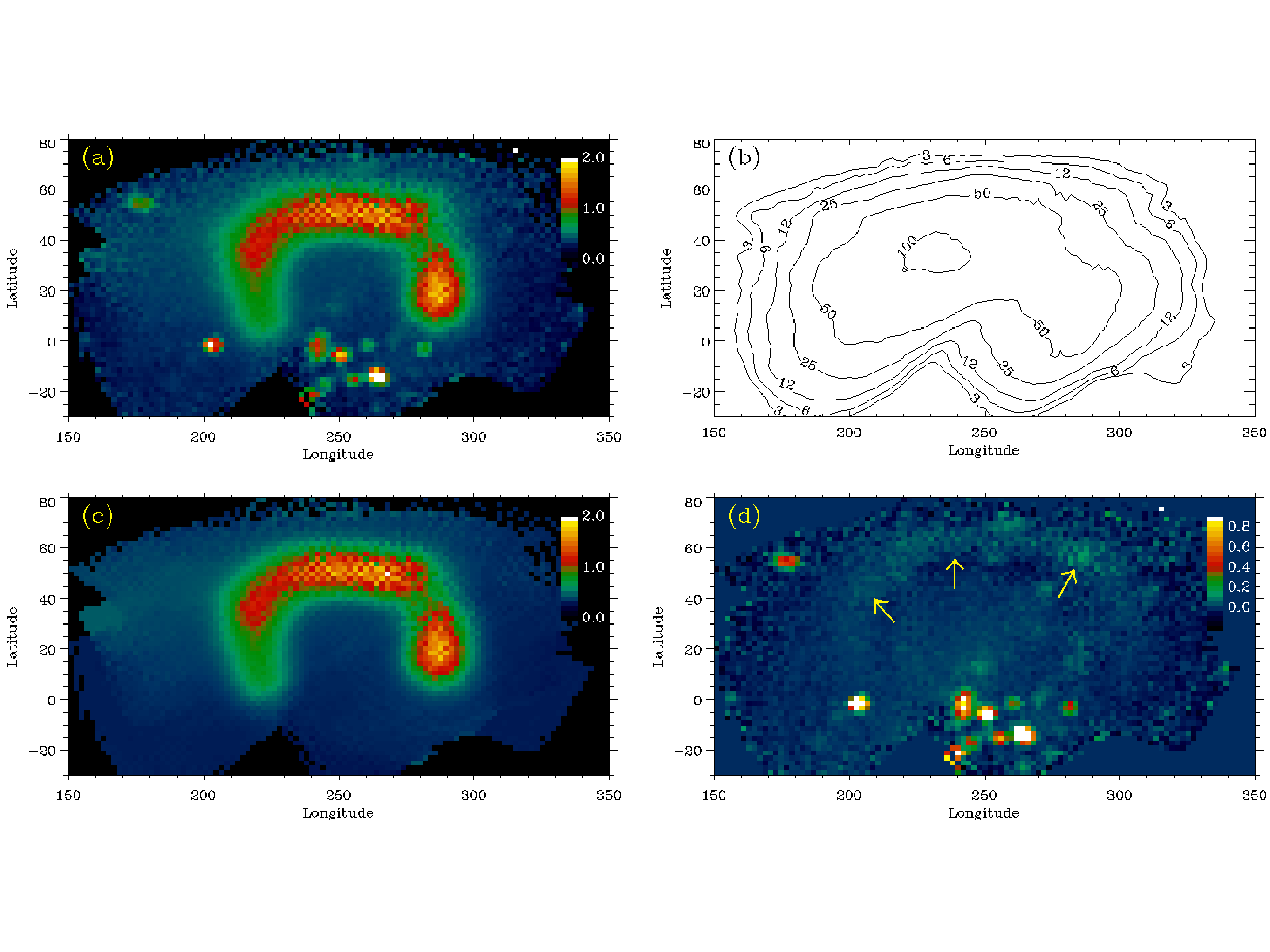}{3.6in}{0}{70}{70}{-255}{-57}
\caption{(a) Map of average count rates observed by {\em Ulysses}/GAS for
  238 WFOV He beam maps observed throughout the mission, in ecliptic
  coordinates.  Point sources are UV-bright stars.  The horseshoe-shaped
  streak is the track of the He beam across the sky during the course of
  {\em Ulysses}'s orbit around the Sun. (b) Contour plot indicating the
  number of individual maps that actually sample each grid point.
  (c) Reconstruction of the coadded count rate map in (a) based on the
  best-fit He flow parameters and background values from WMW15. (d) Residual
  map after subtracting (c) from (a).  Yellow arrows identify the
  residual signal that is interpreted as being from secondary He neutrals.}
\end{figure}
     Thus, in searching for the secondary signal, we first coadd all
238 beam maps.  The individual maps are irregularly gridded in
ecliptic coordinates, but we map them onto a regular grid,
with a grid point size of $2^{\circ}\times 2^{\circ}$.
At each grid point, $i$, we determine the average count rate observed
at that point for the set of beam maps that actually sample that
location, $S_i$.  By keeping track of the effective exposure time for
each bin, $t_i$, we can also compute the Poissonian uncertainties
of the $S_i$ count rates, $\sigma_i=\sqrt{S_it_i}/t_i$.
The resulting map of $S_i$ count rates is shown in Figure~1(a).  The
point sources in the map are UV-bright stars, as the GAS instrument
possesses some degree of UV sensitivity.  But the primary signal
apparent in the image is a horseshoe-shaped feature, which represents
the track of the primary He beam across the sky, as {\em Ulysses}'s
position and motion vector change during the course of its orbit.  When
the ISM He atoms become visible at
the beginning of a fast latitude scan, with {\em Ulysses} south of the
ecliptic plane, the beam is observed at the right end of the
horseshoe, near ($\lambda$,$\beta$)=($285^{\circ}$,$10^{\circ}$).  The
beam then shifts upwards to about
($280^{\circ}$,$45^{\circ}$) at the ecliptic plane crossing.  With
{\em Ulysses} moving north of the ecliptic and ultimately away from
the Sun, the beam then shifts to the left and then ultimately
downwards, ending up at the left end of the horseshoe near
($220^{\circ}$,$10^{\circ}$), when the He atoms become unobservable again.

     In interpreting Figure~1(a), it is worth noting again that each
grid point is the average count rate observed at that point for {\em all}
maps that include that point, including maps where that point is actually
outside the location of the beam at that time.  Furthermore, we are
mapping irregularly gridded points onto a regular grid, so adjacent
points can actually be sampled by a different fraction
of on-beam to off-beam maps.  This is the primary reason for the
pixel-to-pixel variation within the beam, not Poissonian noise.
Likewise, the intensity variation along the horseshoe-shaped beam is
mostly an indication of different parts of the horseshoe being sampled
by a different number and fraction of on-beam maps.  For example,
the gap in the horseshoe at about ($280^{\circ}$,$40^{\circ}$) is a
location where there are relatively few on-beam maps (see Figure~3(a)
in WMW15), so the available count rate measurements for that
location include mostly off-beam measurements.  Thus, the average
count rate there ends up low.  Figure~1(b)
is a contour plot showing the number of maps sampled at each point
in the grid.  Within the horseshoe-shaped track, each point is
typically sampled by $50-100$ individual maps, out of the 238 total
considered.  The middle of the horseshoe is similarly well sampled, but
the sampling falls off quickly outside the horseshoe.

     In searching for a secondary signal, the first step is to subtract
the primary beam from the data.  In WMW15, we performed a global fit to
the primary He neutrals observed in the 238 maps, deriving
a best-fit He flow vector, and best-fit synthetic beam maps.  Each of
these individual maps assumes a flat background underneath the beam,
and these backgrounds are free paramters of the global fit.
We can coadd the synthetic He beam and background maps in the same way
as we coadded the actual beam maps to yield maps of primary beam count
rates, $C_{1,i}$, and model background, $B_i$.
The sum, $C_{1,i}+B_i$, is shown in Figure~1(c).  Figure~1(d) shows the
residual after subtracting Figure~1(c) from Figure~1(a), 
$S_{2,i}=S_i-C_{1,i}-B_i$.  This is now a map in which we can actually
search for a signal from the secondary He flow.

     There does seem to be an excess signal after the subtraction
of the primary beam, identified by arrows in Figure~1(d).  We claim this
to be a likely {\em Ulysses} detection of the ``Warm Breeze''
neutrals first observed by IBEX.  In their fit of the neutrals,
\citet{mak16} find a much slower flow speed of $V=11.3$ km~s$^{-1}$
compared to the $V=26.08$ km~s$^{-1}$ flow of the primary He component
(WMW15).  In {\em Ulysses} data, a slower flow should lead to a
significantly larger horseshoe-shaped track than that of the primary
He beam (see Figure~7 in WMW15).  Thus, based on the \citet{mak16}
fit, the expectation is that in Figure~1 the secondary He component
should be visible as a faint large horseshoe surrounding the bright
smaller horseshoe of the primary component.  This is a reasonable
description of the residual signal seen in Figure~1(d), though the
``legs'' of the horseshoe are not as visible.  It is also worth
noting that the set of images that defines the residual signal at one
location within the horseshoe will be very different from the set of
images that defines the signal at a very different location within
the horseshoe.  It is impressive that a coherent horseshoe shape in
Figure~1(d) emerges from beam maps that individually contribute to
only one part of the horseshoe.

\section{Fitting the {\em Ulysses} Secondary Helium Component}

     We now fit the {\em Ulysses} secondary He component, using
techniques analogous to those used previously to fit the primary
component, assuming a laminar, Maxwellian flow from infinity (WMW15).
We have already noted in Section~1 that the laminar flow assumption
is a poor one for the secondary He, but there are two reasons for
keeping it for now.  The first is simplicity, as it allows the
secondary signal to be modeled using the same codes used to analyze
the primary He beam.  The second reason is that we want to be able
to compare our {\em Ulysses} results with IBEX, and the IBEX secondary
He flow properties are currently inferred assuming a laminar flow
\citep{mak16}.

     The five fit parameters are flow speed ($V$), flow longitude ($\lambda$)
flow latitude($\beta$), temperature ($T$), and density far from the Sun
($n_{\rm He}$).  One difference is that because it is very difficult to
visually see the secondary signal in individual beam maps, we fit the
coadded residual count rate map, $S_{2,i}$, rather than the
direct {\em Ulysses} measurements in the individual maps,
which was the approach in fitting the primary beam.
We refer the reader to Section~4 of WMW15 for details
about the particle tracking approach and synthetic map generation.
After the 238 synthetic beam maps are created, they are coadded as the
observed maps were coadded in Figure~1, in order to compare with
the residual signal in Figure~1(d).  The best fit is determined by
minimizing the $\chi^2$ statistic \citep{prb92}.  The average
1~AU photoionization rate for the 238 {\em Ulysses} maps under
consideration is $\beta_{ion}=1.5\times 10^{-7}$ s$^{-1}$ (WMW15, see
Section~5), so we simply assume that value in our calculations.

     The WMW15 analysis did not consider the contribution
of the secondary neutrals to the background, and therefore the
true background may have been overestimated, meaning Figure~1(d)
could be underestimating the secondary signal.  In order
to correct for this, we first perform a preliminary fit to the
secondary signal.  We then use the best-fit secondary He flow
parameters to compute the secondary count rates within the 238
original {\em Ulysses} beam maps.  Subtracting the secondary counts
from these maps, we then redo the WMW15 analysis of the primary beam.
This accomplishes two things.  The first is to see
whether correcting for the secondaries in any way affects the fit
parameters of the primary He flow.  The answer is that it
does not.  The secondary signal is too weak to significantly affect
the fit to the primary neutrals.  The second accomplishment is revised
measurements of the background levels in the individual 238
maps, which we can use to create a revised $B_i$ background map.
The average background of the 238 maps is 0.384 cts~s$^{-1}$,
which is not that different from the original measurements
(see Figure~8(b) in WMW15).

\begin{figure}[t]
\plotfiddle{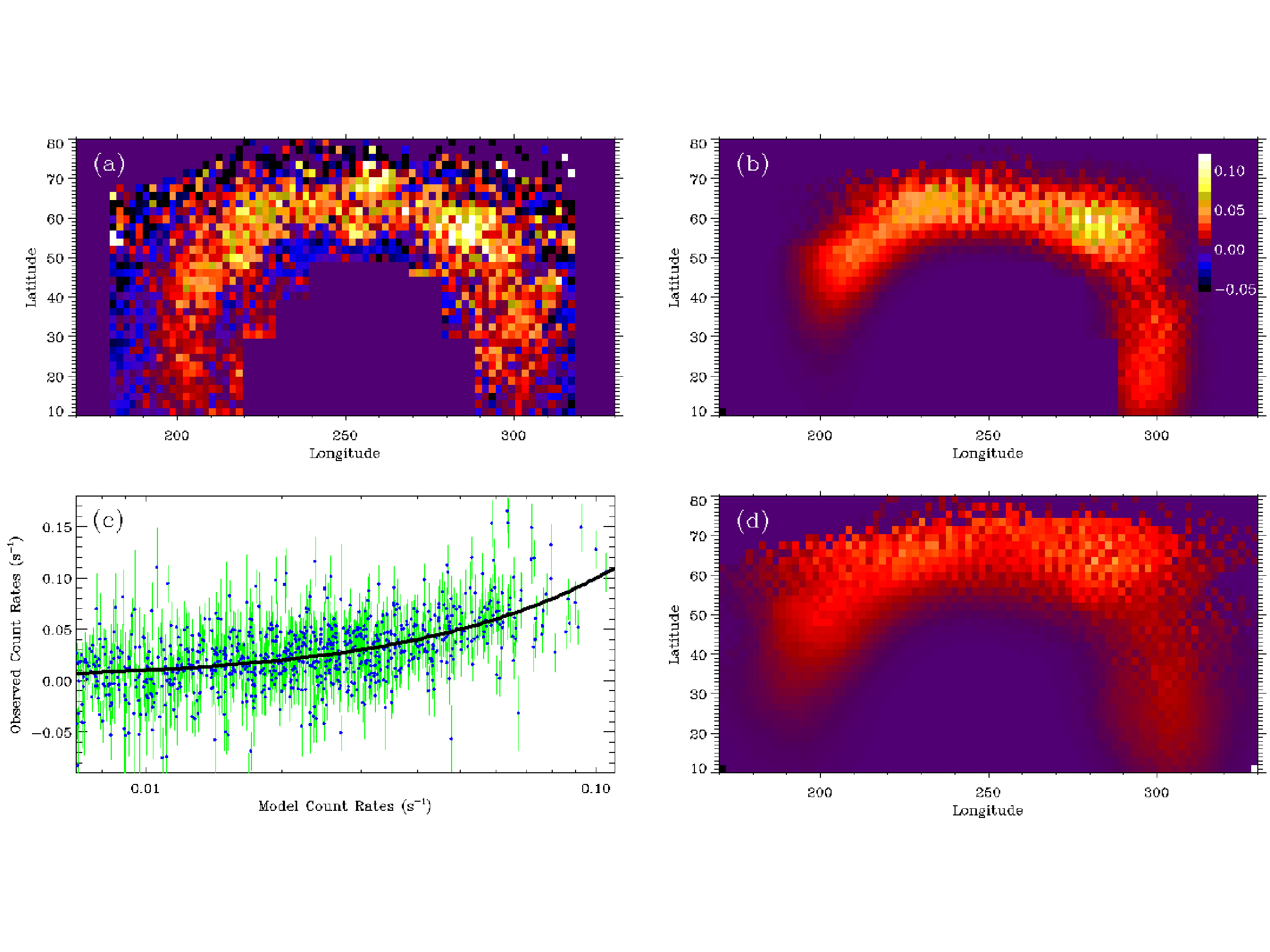}{3.6in}{0}{70}{70}{-255}{-57}
\caption{(a) A map of residual counts after subtracting the primary
  beam and background counts, zooming
  in on only the pixels around the He secondary signal that are being
  fitted.  (b) Average count rate map based on the best He flow vector
  fit to the data in (a).  (c) Observed versus model count rates based
  on the best He flow vector fit, with the thick black line being the
  line of agreement.  A log scale is used for the x-axis to spread out
  the low count values.  (d) Average count rate map expected based on
  the He secondary flow vector inferred from IBEX data (Kubiak et
  al.\ 2016).  Note that the count rate color scale in (b) applies to
  all the count rate map panels.}
\end{figure}
     We recompute the residual map, $S_{2,i}$, using the revised
background estimates.  Figure~2(a) displays the revised $S_{2,i}$ map,
which is not greatly different from that in Figure~1(d).
The figure zooms in on the region around the secondary signal, identifying
the pixels in the map used in our fits.  A new and final fit is
performed to this count rate map.  Figure~2(b) shows the count rate
map of our best fit to the data, which we denote as $C_{2,i}$.
Figure~2(c) explicitly compares
the observed and modeled count rates in this fit.  Typical count rates
within the horseshoe-shaped secondary He beam track are $0.05-0.10$
cts~s$^{-1}$.  This is a factor of 4 to 8 lower than the background
($B_i$) and a factor of 20 to 40 lower than typical count rates within
the primary beam track ($C_{1,i}$), both of which have been subtracted from
the data to yield the residual signal in Figure~2(a).  In addition to
roughly reproducing the shape and width of the horeshoe-shaped track,
it is encouraging that the fit reproduces the bright spot
on this track observed at ($\lambda$,$\beta$)=($285^{\circ}$,$60^{\circ}$),
which has no real analog on the primary beam track (see Figure~1).
There is, therefore, a strong case for this being a real feature.
Finally, Figure~2(d) shows the secondary He signal predicted by the
IBEX-derived ``Warm Breeze'' flow vector described in Section~1.

\begin{figure}[t]
\plotfiddle{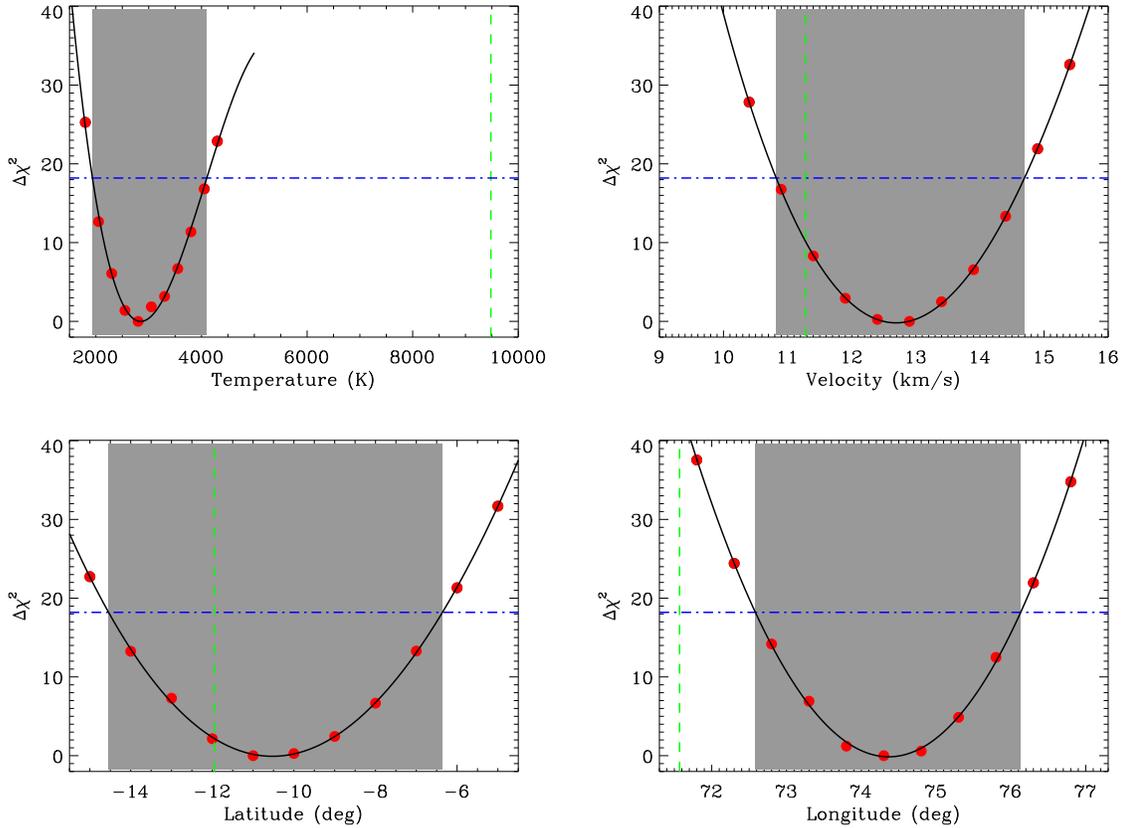}{4.1in}{90}{65}{65}{260}{-45}
\caption{Secondary helium flow parameter measurements from a fit to the
  {\em Ulysses} coadded count rate residual map in Figure~2(a).  In
  each panel, $\Delta \chi^2$ is plotted as a function of one of the four
  He flow parameters of interest, where each point represents a separate
  fit with that parameter held constant and the other three parameters
  (and the He density) allowed to vary freely.  Solid lines show
  polynomial fits to the data points.  The horizontal
  dot-dashed line corresponds to the $3\sigma$ contour used to
  define the uncertainty range in each parameter (shaded regions).
  Vertical dashed lines are the Kubiak et al.\ (2016) measurements
  from IBEX data.}
\end{figure}
     Determining the best fit parameters and their uncertainties requires
computing a grid of fits.  Figure~3 shows how $\chi^2$ varies when $T$,
$V$, $\beta$, and $\lambda$ are held constant.  The $\chi^2$ statistic
is well-behaved, showing a clear $\chi^2$ minimum, $\chi^2_{min}$, in
each panel.  If $\nu$ is the number of degrees of freedom of the fit
(the number of data points minus the number of free parameters), then
the reduced chi-squared is defined as $\chi^2_{\nu}=\chi^2/\nu$, which
should be $\sim 1$ for a good fit.  For our fit, $\nu=1721$ and
$\chi^2_{{\nu},min}=2.397$.  We define $\Delta \chi^2 \equiv
\chi^2-\chi^2_{min}$, with each panel of Figure~3 showing the variation
of $\Delta \chi^2$ across the $\chi^2_{min}$ region.  Third order
polynomials are fitted to the data points to interpolate between them.

     The $\Delta \chi^2$ values are used to define the error bounds around
$\chi^2_{min}$, as described by \citet{prb92} and
\citet{whp89}.  For the number of free parameters of our fit (five),
the $3\sigma$ confidence contour corresponds to $\Delta \chi^2=18.2$,
based on relation 26.4.14 of \citet{ma65}, and
this level defines the uncertainty ranges shown in Figure~3.
Our derived secondary He flow parameters are: $V=12.8\pm 1.9$ km~s$^{-1}$,
$\lambda=74.4\pm 1.8^{\circ}$, $\beta=-10.5\pm 4.1^{\circ}$, and
$T=3000\pm 1100$~K.  At this point it should be noted that
{\em Ulysses} data are provided in B1950 coordinates, and the analysis
is performed in those coordinates, but the coordinates we quote in
this paper are always converted to the now more standard J2000 epoch.

     In Figure~3, the {\em Ulysses}-derived flow parameters are compared
with those from IBEX data, also listed in Section~1.  The $V$ and
$\beta$ values are in good agreement.  There is a small inconsistency
in $\lambda$, but if the total statistical plus systematic
uncertainty in the IBEX measurement ($0.5^{\circ}+0.9^{\circ}=1.4^{\circ}$) is
considered \citep{mak16}, the IBEX and {\em Ulysses} error bars
will overlap.  The only serious disagreement is temperature, $T$, for
which there is a very large discrepancy.  Figure~2(d) shows how the
larger $T$ of the IBEX secondary parameters leads to a broader
horseshoe than observed.  For example, there is significant predicted
flux at ($\lambda$,$\beta$)=($200^{\circ}$,$60^{\circ}$) and
($300^{\circ}$,$70^{\circ}$), which does not seem to be observed
in Figure~2(a).  The relatively narrow observed horseshoe-shaped
track seems to require a surprisingly low temperature of
$T\approx 3000$~K.

     Before discussing the temperature issue further, there is one
final fit parameter to discuss, the density.
For the fits in Figure~3 that fall within the error bars, we compute
the mean and standard deviation of the He densities inferred from
these individual fits, leading to our best estimate of the He density of the
secondary component, $n_{\rm He}=(9.6\pm0.7)\times 10^{-4}$ cm$^{-3}$.
Dividing this by the best estimate of the primary component density from
WMW15, $n_{\rm He}=0.0196$ cm$^{-3}$, we find that the secondary
component is $4.9\pm 0.9$\% of
the primary component, in good agreement with the 5.7\% measurement
from IBEX.  Four of the five {\em Ulysses} secondary He flow
parameters are therefore in reasonably good agreement with IBEX
measurements, providing support for the detection of the ``Warm Breeze''
neutrals by {\em Ulysses}.

     It is only the temperature that seems inconsistent.  The
IBEX-derived $T=9500$~K temperature was already recognized to be
problematic, being well below the $T\approx 20,000$~K temperatures
expected for the outer heliosheath source region of the secondary
He neutrals.  Our {\em Ulysses} measurement of $T=3000$~K represents
an even larger underestimate.  The implausibly low temperature
measurements are a consequence of the improper
approximation of the secondary neutral flow as being laminar beyond
the heliopause, as opposed to a divergent flow due to deflection
around the heliopause.

     This was demonstrated explicitly by
\citet{bew17}, who used a simple 2-D model of a divergent
flow field to explore how the angular width of the observed He beam
observed near 1~AU, $W$, relates to flow velocity, $V$, temperature,
$T$, and flow divergence, $D$, in the outer heliosheath source region.
The flow pattern at the outer boundary is defined by the simple
equation $\psi=D\times \theta$, where $\theta$ is the viewing angle
from the upwind direction of the ISM flow, and $\psi$ is the deviation
of flow direction from the ISM flow direction.  So if $D=0.5$, at
an angular distance of $\theta=30^{\circ}$ from the upwind direction
the flow would be diverging by $\psi=15^{\circ}$ from the direction
of the ISM flow (see Figure~1 of Wood 2017).  It was shown that
$W$, $V$, $T$, and $D$ could be related by the power law relation
\begin{equation}
W=C\left( \frac{V}{20} \right)^{\alpha} \left( \frac{T}{10^4} \right)^{\beta}
  \left( D+1 \right)^{\gamma},
\end{equation}
with $V$ in km~s$^{-1}$ and $T$ in K.  For the case with an observer
at 1~AU along the stagnation axis, $C=24.1^{\circ}$, $\alpha=-0.84$,
$\beta=0.52$, and $\gamma=-0.91$.

     Applying equation~(1), the {\em Ulysses} fit to the
secondaries, ($D$,$T$,$V$)=(0,3000,12.8), would predict a He beam
width of $W=18.7^{\circ}$ in the context of the simple 2-D model.
Global heliospheric models suggest $V\approx 9$ km~s$^{-1}$ and
$T\approx 21,000$~K for the outer heliosheath \citep{mak14,vvi15}.
Assuming these values for $V$ and $T$, equation~(1) can then be used
to compute the value of $D$ necessary to recover the $W=18.7^{\circ}$
width that is crudely representative of the {\em Ulysses}
measurements.  The resulting divergence is $D=3.2$.  The point is that
a sufficiently divergent flow can explain the low temperature
measurement.  Equation~(1) and the values of the power law indices
quoted above will not be precisely applicable to either the
{\em Ulysses} or IBEX cases, as the 2-D model does not accurately
replicate the observing geometries of either, but we use them here
simply to illustrate how switching from a laminar to a divergent flow
should naturally lead to higher and more plausible temperatures.

     This does not necessarily explain why the {\em Ulysses}
temperature measurement is so much lower than the IBEX
measurement.  Is it possible that the {\em Ulysses} data are more
sensitive to the effects of a divergent flow than IBEX?
This could in principle be the case, given the very different
observing geometries of {\em Ulysses} and IBEX, and the energy-dependent
sensitivity of the {\em Ulysses}/GAS detector, which is unlike
the IBEX detector.  Exploring this further would
require fitting the IBEX and {\em Ulysses} data again assuming a
divergent flow rather than a laminar one, to see if such an
assumption would not only lead to higher and more plausible $T$
measurements, but would also resolve the IBEX/{\em Ulysses}
discrepancy.  Such a task is beyond the scope of the present
analysis.

     Another concern particular to the {\em Ulysses} measurement is
the issue of background subtraction.  Discussion of {\em Ulysses}
background sources can be found in \citet{mw93} and \citet{mb96}.
The secondary signal is
significantly weaker than the background, so inaccuracies in the
background subtraction represent a significant source of systematic
uncertainty in the analysis.  Hints
of inaccuracies are apparent in Figure~2(a), where residual count
rates outside the secondary signal seem preferentially negative, when
they should be zero on average.  We compute an average count rate
of $-0.008$ cts~s$^{-1}$ outside the secondary signal in Figure~2(a),
suggesting that our background map, $B_i$, may be $\sim 2$\% too high
in the vicinity of the secondary signal.

     Our analysis of the primary He beam assumes a flat background
under the He beam in each of the 238 individual beam maps.  In order to
explore whether the flat background assumption might be affecting our
results, we redid the analysis allowing the background to vary in a
linear fashion underneath the beam.  However, the resulting
coadded background map, $B_i$, does not end up looking very different
from the coadded map assuming flat backgrounds, and the residual maps
$S_{2,i}$ shown in Figures 1(d) and 2(a) therefore do not look very
different either.  Thus, the assumption of a non-flat background does
not significantly affect the secondary flow fit parameters.

     In a final effort to see if we can improve agreement with the IBEX
measurements, we conduct the following experiment.  We described above
how we revised the background measurements from WMW15 for the individual
238 {\em Ulysses} maps using a preliminary fit to the $S_{2,i}$ residual
map to account for the secondary neutrals.  We repeat this revision,
but we instead use the IBEX He flow parameters to estimate the
secondary neutral count rates, analogous to how Figure~2(d) was
computed.  This leads to a revised measurement of the background
map, $B_i$, and a revised $S_{2,i}$ residual map.  We are essentially
trying to bias the analysis towards ultimately yielding a residual map
that looks more like Figure 2(d) than Figure 2(a).  However, results are
once again not much different than before, and when we fit the resulting
residual map, there is no significant change in the fit parameters.
Thus, this test fails to provide evidence that background uncertainties
are responsible for the $T$ discrepancy with IBEX.

\section{Deflection from the ISM Flow Direction}

     Possibly the strongest evidence that the ``Warm Breeze''
neutrals detected by IBEX are created by charge exchange in the
outer heliosheath is that the flow direction of the ``Warm Breeze''
neutrals lies between the ISM flow direction and the direction of
the ISM magnetic field.  The idea is that the ISM field creates
asymmetries in the heliopause and the flow around the heliopause,
such that neutrals created by charge exchange outside the heliopause
will appear to be deflected from the ISM flow direction towards the
direction of the ISM magnetic field \citep{vi05,mo07,nvp08}.

\begin{figure}[t]
\plotfiddle{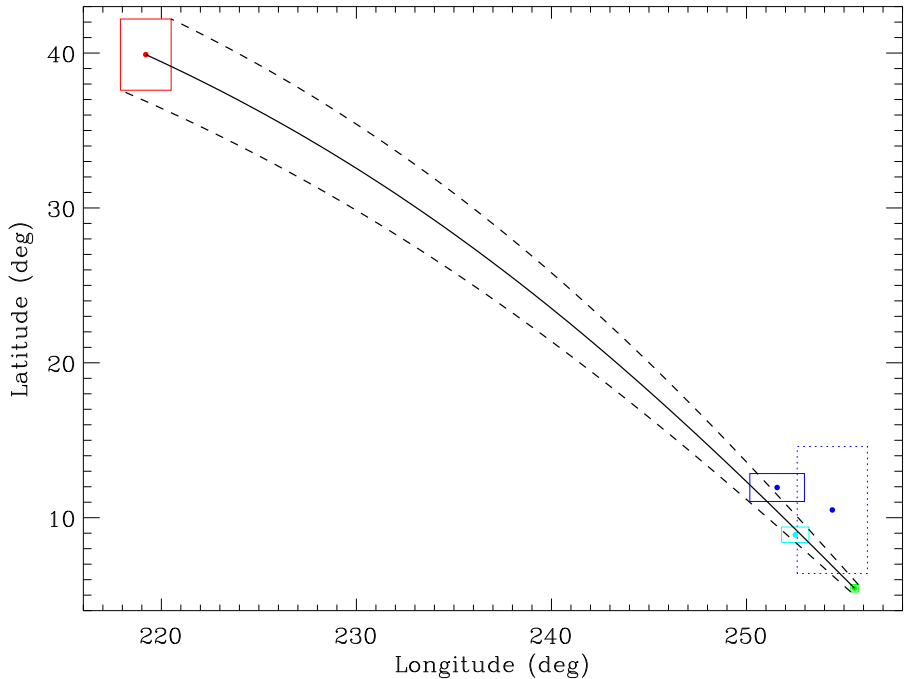}{3.7in}{90}{55}{55}{200}{-40}
\caption{A map in ecliptic coordinates showing the assumed direction
  of the local ISM magnetic field based on the center of the IBEX
  ribbon (red box; Funsten et al.\ 2013), the upwind direction of the
  ISM He flow (green box; WMW15), the upwind direction of the H flow
  in the solar system (light blue box; Lallement et al.\ 2005, 2010),
  the IBEX-derived upwind direction of the secondary He flow
  (solid blue box; Kubiak et al.\ 2016), and the {\em Ulysses}-derived
  upwind direction of the secondary He flow (dotted blue box; this
  paper).  The boxes indicate the quoted uncertainties in the
  directions.  The ISM velocity and field directions define a plane
  whose intersection with the map is shown as a great circle line
  connecting the velocity and field directions, with uncertainties
  indicated by dashed lines.  The expectation is
  that the H and secondary He flows should be on this line.}
\end{figure}
     In Figure~4, we plot in ecliptic coordinates the ISM field
direction inferred from the IBEX ribbon center \citep{hof13}
and the upwind ISM flow direction from WMW15, with the connecting line
indicating the path along a great circle connecting the two.
The secondary He neutral flow direction should lie on this line.
The He secondary flow directions from both IBEX \citep{mak16} and
{\em Ulysses} are shown, and both IBEX and {\em Ulysses} boxes overlap
the line, albeit in different locations.  Likewise, the
expectation is that the H flow direction
measured by the SWAN instrument on SOHO should also lie along this
line, as the observed flow should include many H neutrals created by
charge exchange beyond the heliopause.  The SOHO/SWAN measurements of
the H flow are shown in Figure~4 \citep{rl05,rl10}, demonstrating that
the H flow does indeed lie nicely on the line.

     The IBEX-derived flow direction has lower uncertainties and
is closer to the expected line of deflection than our {\em Ulysses}
measurement.  For the primary ISM He flow, we actually found
that {\em Ulysses} constraints on the He flow vector are
tighter than those of IBEX despite lower S/N \citep[WMW15,][]{bew15a}.
This is due to the {\em Ulysses} advantage
of observing the He flow at different locations and with different
spacecraft motion vectors, which effectively breaks the parameter
degeneracies that plague analysis of IBEX data.  However, for the fainter
secondary component, it is likely that the superior S/N of IBEX
is more important and will lead to tighter constraints on the secondary
flow vector.  Still, the {\em Ulysses} data provide a useful
independent confirmation of the existence and general characteristics
of the secondary He flow in the inner heliosphere.

\section{Summary}

     We have analyzed {\em Ulysses}/GAS observations of neutral He
flowing through the solar systeam to search for a signature
of the ``Warm Breeze'' neutrals detected by IBEX, and our results
are summarized as follows:
\begin{enumerate}
\item By coadding the {\em Ulysses} He maps together to maximize S/N,
  and then subtracting the signal from the primary He neutrals and the
  background, we are able to find a weak residual signal that
  represents a likely detection of the secondary He neutrals first
  detected by IBEX.
\item We estimate the secondary He flow vector by fitting the
  residual signal assuming a laminar flow at infinity,
  yielding the following fit parameters:  $V=12.8\pm 1.9$ km~s$^{-1}$,
  $\lambda=74.4\pm 1.8^{\circ}$, $\beta=-10.5\pm 4.1^{\circ}$, and
  $T=3000\pm 1100$~K; with a density $4.9\pm 0.9$\% that of the primary
  ISM He component.  Most of these values are in reasonable
  agreement with IBEX measurements \citep{mak16}, which were
  also based on the laminar flow at infinity approximation.  The one
  discrepant parameter is temperature, where our value is much lower
  than the IBEX-derived $T=9500$~K.  Both the IBEX and {\em Ulysses}
  temperatures are significantly lower
  than those expected to actually exist in the outer heliosheath.
  The underestimates most probably will be due to the assumption of a
  laminar flow at the outer boundary, as opposed to the divergent flow
  that will actually exist there.
\item The reason for the discrepant $T$ measurement is uncertain at
  this time, but it could also be due to background subtraction
  uncertainties for {\em Ulysses} \citep{mw93,mb96}, or it could be
  associated with the inadequate and imprecise assumption that the He
  secondaries can be approximated as a laminar flow from infinity,
  which could in principle affect the IBEX and {\em Ulysses} data
  differently.
\end{enumerate}

     The He secondary flow is a unique diagnostic of the outer
heliosheath that may prove to be the best way to remotely study the
deflection of the ISM flow around the heliopause.  The biggest problem
with existing analyses of the secondary He is the approximation of
the flow as a laminar flow from infinity.  Future analyses should
instead assume a parametrized divergent flow from about 150~AU.  This
should lead to more accurate inferences of the flow properties in the
outer heliosheath, including a higher and more realistic
temperature.  It remains to be seen if such an analysis will resolve
the $T$ discrepancy between IBEX and {\em Ulysses} reported here.

\acknowledgments

Support for this project was provided by NASA award NNH16AC40I to
the Naval Research Laboratory.


\begin{thebibliography}{}

\bibitem[Abramowitz \& Stegun(1965)]{ma65}
Abramowitz, M., \& Stegun, I. A. 1965, Handbook of Mathematical Functions
  (New York:  Dover Publications, Inc.)
\bibitem[Banaszkiewicz et al.(1996)]{mb96}
Banaszkiewicz, M., Witte, M., \& Rosenbauer, H. 1996, A\&AS, 120, 587
\bibitem[Bevington \& Robinson(1992)]{prb92}
Bevington, P. R., \& Robinson, D. K. 1992, Data Reduction and Error
  Analysis for the Physical Sciences (New York: McGraw-Hill)
\bibitem[Bzowski et al.(2017)]{mb17}
Bzowski, M., Kubiak, M. A., Czechowski, A., \& Grygorczuk, J. 2017,
  ApJ, 845, 15
\bibitem[Bzowski et al.(2014)]{mb14}
Bzowski, M., Kubiak, M. A., Hlond, M., et al. 2014, A\&A, 569, A8
\bibitem[Bzowski et al.(2012)]{mb12}
Bzowski, M., Kubiak, M. A., M\"{o}bius, E., et al. 2012, ApJS, 198, 12
\bibitem[Frisch et al.(2013)]{pcf13}
Frisch, P. C., et al. 2013, Science, 341, 1080
\bibitem[Funsten et al.(2013)]{hof13}
Funsten, H. O., DeMajistre, R., Frisch, P. C., et al. 2013, ApJ, 776, 30
\bibitem[Izmodenov \& Alexashov(2015)]{vvi15}
Izmodenov, V. V., \& Alexashov, D. B. 2015, ApJS, 220, 32
\bibitem[Izmodenov et al.(2005)]{vi05}
Izmodenov, V., Alexashov, D., \& Myasnikov, A. 2005, A\&A, 437, L35
\bibitem[Izmodenov et al. (2003)]{vi03}
Izmodenov, V., Malama, Y. G., Gloeckler, G., \& Geiss, J. 2003, ApJ, 594, L59
\bibitem[Katushkina et al.(2014)]{oak14}
Katushkina, O. A., Izmodenov, V. V., Wood, B. E., \& McMullin, D. R. 2014,
  ApJ, 789, 80
\bibitem[Kubiak et al.(2014)]{mak14}
Kubiak, M. A., Bzowski, M., Sok\'{o}l, J. M., et al. 2014, ApJS, 213, 29
\bibitem[Kubiak et al.(2016)]{mak16}
Kubiak, M. A., Swaczyna, P., Bzowski, M., et al. 2016, ApJS, 223, 25
\bibitem[Lallement et al.(2005)]{rl05}
Lallement, R., Qu\'{e}merais, E., Bertaux, J. L., et al. 2005, Science, 307,
  1447
\bibitem[Lallement et al.(2010)]{rl10}
Lallement, R., Qu\'{e}merais, E., Koutroumpa, D., et al. 2010, in {\em
  Solar Wind 12}, ed. M. Maksimovic et al. (Melville, NY: AIP), 555
\bibitem[McComas et al.(2009)]{djm09}
McComas, D. J., Allegrini, G., Bochsler, P., et al. 2009,
  Space~Sci.~Rev., 146, 11
\bibitem[McComas et al.(2015)]{djm15}
McComas, D. J., Bzowski, M., Frisch, P., et al. 2015, ApJ, 801, 28
\bibitem[M\"{o}bius et al.(2012)]{em12}
M\"{o}bius, E., Bochsler, P., Bzowski, M., et al. 2012, ApJS, 198, 11
\bibitem[M\"{u}ller et al.(2013)]{hrm13}
M\"{u}ller, H. -R., Bzowski, M., M\"{o}bius, E., \& Zank, G. P.
  2013, in {\em Solar Wind 13}, ed. G. P. Zank, et al. (New York:
  AIP, Vol.\ 1539), 348
\bibitem[Opher et al.(2007)]{mo07}
Opher, M., Stone, E. C., \& Gombosi, T. I. 2007, Science, 316, 875
\bibitem[Park et al.(2016)]{jp16}
Park, J., Kucharek, H., M\"{o}bius, E., et al. 2016, ApJ, 833, 130
\bibitem[Pogorelov et al.(2008)]{nvp08}
Pogorelov, N. V., Heerikhuisen, J., \& Zank, G. P. 2008, ApJ, 675, L41
\bibitem[Press et al.(1989)]{whp89}
Press, W. H., Flannery, B. P., Teukolsky, S. A., \& Vetterling, W. T.
  1989, Numerical Recipes (Cambridge: Cambridge University Press)
\bibitem[Schwadron et al.(2016)]{nas16}
Schwadron, N. A., M\"{o}bius, E., McComas, D. J., et al. 2016, ApJ, 828, 81
\bibitem[Sok\'{o}{\l}~et al.(2015)]{jms15}
Sok\'{o}{\l}, J .M., Kubiak, M. A., Bzowski, M., \& Swaczyna, P. 2015, ApJS,
  220, 27
\bibitem[Wenzel et al.(1992)]{kpw92}
Wenzel, K. -P., Marsden, R. G., Page, D. E., \& Smith, E. J. 1992, A\&AS,
  92, 207
\bibitem[Witte(2004)]{mw04}
Witte, M. 2004, A\&A, 426, 835
\bibitem[Witte et al.(1996)]{mw96}
Witte, M., Banaszkiewicz, M., \& Rosenbauer, H. 1996, Space Sci.~Rev., 78, 289
\bibitem[Witte et al.(1993)]{mw93}
Witte, M., Rosenbauer, H., Banaszkiewicz, M., \& Fahr, H. 1993,
  Adv.~Space~Res., 13, 121
\bibitem[Witte et al.(1992)]{mw92}
Witte, M., Rosenbauer, H., Keppler, E., Fahr, H., Hemmerich, P., Lauche, H.,
  Loidl, A., \& Zwick, R. 1992, A\&AS, 92, 333
\bibitem[Wood(2017)]{bew17}
Wood, B. E. 2017, in {\em 16th Annual International Astrophysics Conference:
  Turbulence, Structures, and Particle Acceleration Throughout the Heliosphere
  and Beyond}, J. of Phys. Conf. Ser., 900, 012021
\bibitem[Wood \& M\"{u}ller(2015)]{bew15a}
Wood, B. E, \& M\"{u}ller, H. -R. 2015, in {\em 14th Annual International
  Astrophysics Conference:  Linear and Nonlinear Particle Energization
  Throughout the Heliosphere and Beyond}, J. of Phys. Conf. Ser., 642, 012029
\bibitem[Wood et al.(2015)]{bew15b}
Wood, B. E., M\"{u}ller, H. -R., \& Witte, M. 2015, ApJ, 801, 62 (WMW15)
\bibitem[Zank et al.(2013)]{gpz13}
Zank, G. P., Heerikhuisen, J., Wood, B. E., et al. 2013, ApJ, 763, 20

\end{thebibliography}
\end{document}